\documentclass[aps,twocolumn,twoside,nofootinbib,prl,preprintnumbers]{revtex4}

\newcommand{\gsim}{\gtrsim}

\newcommand{\Mpl}{M_{\rm Pl}}

\newcommand{\Mbh}{M_{\rm bh}}

\begin{document}
\bibliographystyle{revtex}

\preprint{hep-ph/0207003}

\title{Black hole evaporation with separated fermions}

\author{Tao Han, Graham D. Kribs, and Bob McElrath}
\affiliation{\mbox{Department of Physics, University of Wisconsin, 
Madison, WI 53706} \\
\\
\mbox{than@pheno.physics.wisc.edu, kribs@physics.wisc.edu, 
rsmcelrath@wisc.edu}}


\begin{abstract}
In models with a low quantum gravity scale, 
a well-motivated reason to expect quark and lepton fields are 
localized but physically separated is to avoid proton decay.  
This could happen in a ``fat-brane'' or in an 
additional, orthogonal 1/TeV sized dimension in which the gauge 
and Higgs fields live throughout.  Black holes with masses of order
the quantum gravity scale are therefore expected to evaporate 
non-universally, preferentially radiating directly into quarks or 
leptons but not both.  Should black holes be copiously 
produced at a future hadron collider, we find the ratio of final state 
jets to charged leptons to photons is 113:8:1, which differs 
from previous analyses that assumed all standard model fields live at 
the same point in the extra dimensional space.
\end{abstract}


\maketitle


Black holes are the most captivating prediction of
general relativity despite the dearth of experimental evidence to date.
Testing black hole creation and evaporation in particle 
physics experiments has been considered, until recently, far out-of-reach 
due to the feebleness of gravitational interactions when compared with 
standard model forces.  If space has many compact extra dimensions,
however, the fundamental quantum gravity scale $M_*$ might be as low as 
a TeV \cite{ADD}.  Should we be so lucky, future high 
energy colliders \cite{GT,DL} and ultra-high energy cosmic-ray collisions 
with the atmosphere and the earth's crust \cite{cosmicray} 
can be production sources of 
black holes.  The black holes then quickly decay through Hawking 
radiation \cite{Hawking} on collider time scales, emitting 
energetic jets, leptons, photons, and neutrinos.

Previous analyses have generally assumed that baryon number $B$
and lepton number $L$, or more precisely $B-L$, is conserved
in the evaporation spectrum.
The conservation of $B-L$ must hold to extremely high accuracy
to prevent the proton from decaying through operators suppressed by
the lowered quantum gravity scale, of order a TeV.
However, in large extra dimension scenarios there is no reason for 
$B-L$ to be conserved; additional assumptions must augment 
the absence of a desert.  One approach is to use 
discrete gauge symmetries \cite{KW,discrete}, analogous to imposing 
matter parity in low energy supersymmetric models.  
Alternatively, there is an intrinsically extra dimensional proposal that
physically separates the quark and lepton fields far enough so that their 
wavefunction overlap is exponentially suppressed \cite{AS}.  

Consider a very massive
black hole, with a mass much larger than $M_*$.  Initially, it
is so large as to ``fill up'' all compact extra dimensions.
As the black hole evaporation proceeds, the mass decreases,
the temperature rises, and the horizon shrinks.  Eventually, the 
black hole horizon will become smaller than the large
extra dimensions in which only gravity lives (``gravity-only''
dimensions).  At this point the black hole is radiating mostly
into brane modes, for reasons that were first clearly elucidated 
in \cite{EHM}.  Roughly speaking, the naive enhancement that one
might guess of radiation into a tower of Kaluza-Klein (KK) modes is 
compensated for by the small overlap of the higher dimensional field's 
wavefunction with the black hole horizon.  More detailed calculations 
confirm this intuition \cite{EHM}, which we apply to our scenario.

Since there are many more 
standard model degrees of freedom on the brane over the pristine 
gravity-only bulk, the overwhelmingly dominant evaporation products
are brane modes.  Once the black hole horizon shrinks to become 
smaller than the separation between quarks and leptons, of order
10 to 100 times the inverse quantum gravity scale, a qualitative change is
expected in the black hole evaporation process.  Black holes stuck 
to the brane where quarks live will evaporate directly into quarks but 
not leptons, while black holes stuck to the brane in which leptons live 
will evaporate directly to leptons but not quarks.  This observation has 
dramatic consequences for collider and cosmic-ray signals of black holes.

In a scenario with large extra dimensions, spacetime consists of
our ordinary three space plus one time dimensions plus several 
additional curled-up space dimensions compactified in a volume $V_D$.  
For a torus, for example, $V_D = (2 \pi R)^{D-4}$.  The higher 
dimensional Newton's constant $G_D$ is related to the four dimensional 
one $G_4$ through $G_D = G_4 V_D$.
The coupling of a 4-dimensional graviton to matter is suppressed
by the reduced Planck scale 
\begin{equation}
\Mpl \equiv \frac{1}{\sqrt{8 \pi G_4}} \simeq 2 \times 10^{18}\; {\rm GeV} \; ,
\end{equation}
which is the scale where quantum gravity effects are expected 
to be order one.  The coupling of a higher dimensional graviton to 
higher dimensional matter is analogously suppressed by
\begin{equation}
M_{*} \equiv \left( \frac{1}{8 \pi G_D} \right)^{1/(D-2)}
\end{equation}
which we take to be our definition of the quantum gravity scale.
In a moment it will be clear that this definition is physically
well-motivated using considerations of black hole dynamics.  

The basic properties of black holes in higher dimensions are 
by now well understood.  We will be exclusively concerned with black
holes near the end of their life when their mass $\Mbh$ is larger
than $M_*$ but their horizon $r_h$ is smaller than the size of any
compact dimension.  The reasons for this are twofold:
It is in this regime that the evaporation is expected to be non-universal, 
and secondly future collider or cosmic-ray production 
is expected to create black holes whose mass is not too much larger
than $M_*$.  In all analyses below we treat black holes 
as semi-classical $D$-dimensional objects.  We briefly outline their
salient properties.
The horizon of a $D$-dimensional black hole is given by \cite{MyersPerry}
\begin{equation}
r_h = \frac{1}{M_*} \left( \frac{\Mbh}{M_*} \right)^{1/(D-3)}
\left( \frac{2}{(D-2) \Omega_{D-2}} \right)^{1/(D-3)} \label{r-eq}
\end{equation}
where 
\begin{equation}
\Omega_{D} \equiv \frac{2 \pi^{(D+1)/2}}{\Gamma[(D+1)/2]}
\end{equation}
is the volume of a unit $D$-sphere.  The $D$-dependent numerical 
factor on the far right-hand side of (\ref{r-eq}) varies from 
about $1/4$ to $1/2$ for $6 \le D \le 11$.  
The black hole temperature is simply
\begin{equation}
T = \frac{D-3}{4 \pi r_h}
\end{equation}
that is well approximated (to within $\pm15\%$) by
\begin{equation}
T \simeq M_* \left( \frac{\Mbh}{M_*} \right)^{-1/(D-3)}
\end{equation}
for any number of dimensions $4 \le D \le 11$.  Interestingly,
in four dimensions the temperature is \emph{exactly} equal to the 
reduced Planck scale $T = \Mpl$ for a black hole with mass 
$\Mbh = \Mpl$.  Similarly, with our definition of $M_*$, a higher 
dimensional black hole has a temperature almost identical to $M_*$
for a black hole with mass $\Mbh = M_*$.
No other choice of definition of the higher dimensional Planck scale 
has this property.

We imagine that quarks and leptons are 
brane-localized fields with the smallest physically reasonable 
Gaussian width $\sim$$1/M_*$.  Quarks live on a ``quark brane'', 
while leptons live on a ``lepton brane'', physically separated by
a distance $L \sim (10$--$100)/M_*$, where the uncertainty in
the separation length depends on the nature of the induced proton 
decay operators.  We shall optimistically assume $L \sim 10/M_*$ 
is sufficient to suppress dimension-6 proton decay operators such
as $q q q l/M_*^2$, which should
be correct upon integrating out all but the most bizarre 
(wormhole-like) field configurations.  Gauge and Higgs fields must propagate 
in the bulk 
so that quarks and leptons interact with one another just as in the 
Standard Model.  Of course gauge and Higgs fields will also have a 
KK tower of excitations with masses $M_{KK} = L^{-1}, 
2 L^{-1}$, 
etc., with couplings that may differ significantly
from one-brane-localized models \cite{AGS}. 
In extra dimensional models with bulk gauge and 
Higgs fields but localized matter fermions, there are strong 
experimental constraints on the mass of the lightest KK gauge boson 
from precision electroweak measurements 
(see, e.g., \cite{gaugebulklimits}).  
Although no analysis has been done for this particular scenario,
it is quite reasonable to expect bounds similar to those found for all matter
stuck to the same point, namely $M_{KK} \gsim 2$--$5$ TeV.   
This means that the scale of quantum gravity $M_* \sim 20$--$50$ TeV.

This is a somewhat large scale requiring fine-tuning of the
Higgs mass to one part in about a thousand.  However, a larger $M_*$ 
has several well-known advantages.
Generic four-fermi operators suppressed by 1/TeV give rise to large 
contributions to flavor-changing neutral currents, CP violating 
phenomena, etc., that \emph{cannot} be forbidden by discrete (gauged) 
symmetries.  Here, with $M_*$ in the tens of TeV region, these dangerous
operators are naturally suppressed with order one coefficients.
For example, a recent analysis of the constraint arising 
from neutron--anti-neutron oscillations suggests $M_* \gsim 45$ TeV 
\cite{NSnnbar}, roughly in alignment with the scale in our scenario.
Another advantage of a larger scale for quantum gravity is that it 
allows one fewer gravity-only extra dimensions.  The constraints on 
two extra dimensions from Supernova cooling \cite{constraints} suggest 
$M_*$ should be larger than tens of TeV, but this is automatically 
satisfied in our setup.
We are therefore content with two or more gravity-only extra dimensions
that open up at long distances (up to $0.1$ mm or so), and one 
extra dimension with gauge and Higgs fields propagating in the bulk 
to open up around a few TeV.  The resulting volume of compactified 
space is arranged such that $M_*$ is of order $20$--$50$ TeV.

Black hole formation can occur astrophysically, such as 
from primordial density fluctuations in the early universe, 
through high-energy cosmic-ray collisions, 
or at a sufficiently high energy collider.
There are by now several arguments suggesting black hole
formation can occur at colliders \cite{GT,DL,bhproduction}
(but for opposing views see \cite{voloshin}).  The arguments in favor
of black hole formation are persuasive, but we emphasize that 
the evaporation spectrum is modified near the end of a black hole's
life no matter how it was formed.

The decay of a thermalized black hole proceeds through Hawking 
radiation \cite{Hawking}.  The evaporation rate of a particle species 
$i$ of spin $s$ is given by the usual black-body spectrum
\begin{equation}
\frac{d N_i}{d t} = \frac{c_i \sigma_s}{e^{E/T} - (-1)^{2 s}} 
\frac{d^{n-1} k}{(2 \pi)^{n-1}} \; .
\end{equation}
where $k$ is the ($n-1$)-momenta of the particle living in $n$ dimensions, 
$c_i$ is the multiplicity of the species, and $\sigma_s$ is the absorption
cross section often simply referred to as the
greybody factor.  Note that the number of dimensions, $n$, in which a 
particular standard model field lives should not be confused with 
dimensionality of spacetime $D$.  We find it convenient to rewrite 
the greybody factor as
a dimensionless constant normalized to the black hole surface 
area $A_n$ seen by the $n$-dimensional fields,
\begin{equation}
\Gamma_s = \sigma_s/A_n \; .
\end{equation}
For four-dimensional fields, the emission rate is simply \cite{Page}
\begin{equation}
\frac{d N_i}{d E d t} = \frac{A_4}{8 \pi^2} 
\frac{c_i \Gamma_s E^2}{e^{E/T} - (-1)^{2 s}} \; .
\end{equation}

The greybody factor $\Gamma_s$ is in general both spin- and 
energy-dependent.  Greybody
factors were first calculated by Page \cite{Page} for four-dimensional
black holes.  Higher dimensional greybody factors have not been calculated 
except for scalars \cite{KMR}.  As was emphasized in \cite{Page,MW}, 
one needs the full energy dependence of the greybody factor to calculate 
the full emission spectrum.  However, the integrated power emission 
is reasonably
well approximated by taking the high energy limit of $\Gamma_s$, the
geometric optics approximation.  A black hole acts as a perfect
absorber of a slightly larger radius, $r_c$, given in $D$ dimensions
by \cite{EHM}
\begin{equation}
r_c = \left( \frac{D-1}{2} \right)^{1/(D-3)} \sqrt{\frac{D-1}{D-3}} \, r_h \; .
\end{equation}
We shall employ this approximation to compute the particle emission
and energy spectra in our setup.  The one final refinement that we 
incorporate is to correct for the differing power into 
spin-0, spin-1/2, and spin-1 modes known in four dimensions by 
integrating the spectra numerically \cite{Page}.  This results 
in a suppression of power into spin-1/2 and spin-1 modes 
parameterized by $\Gamma_s$
\begin{equation}
\Gamma_{s=0} = 1 \qquad 
\Gamma_{s=1/2} \simeq 2/3 \qquad 
\Gamma_{s=1} \simeq 1/4 \; . \label{power-suppression-eq}
\end{equation}
Once higher dimensional 
energy-dependent greybody factors become available, they could be 
easily incorporated here, although we do not expect any qualitative 
changes to our results.

The area of a black hole in the geometric optics approximation is 
\begin{equation}
A_n = \Omega_{n-2} \left( \frac{D-1}{2} \right)^{(n-2)/(D-3)} 
      \left( \frac{D-1}{D-3} \right)^{(n-2)/2} r^{n-2} \; .
\end{equation}
The particle emission spectrum into brane fermions is therefore
\begin{equation}
\frac{d N_i}{d E d t} = \frac{c_i \Gamma_s A_4}{8 \pi^2} 
\frac{E^2}{e^{E/T} + 1} \; .
\end{equation}
and into bulk gauge bosons and Higgs is 
\begin{equation}
\frac{d N_i}{d E d t} = \frac{c_i \Gamma_s A_n \Omega_{n-3}}{(n-2)
(2 \pi)^{n-1}} \frac{E^{n-2}}{e^{E/T} - 1} \; .
\end{equation}

The flux emission spectra can be easily integrated, and we find
\begin{equation}
\frac{d N_i}{d t} = \frac{c_i \Gamma_s \Omega_{n-3} A_n f}{(2 \pi)^{n-1} (n-2)}
                   \left( \frac{D-3}{4 \pi r_h} \right)^{n-2} 
                   \Gamma(n-1) \zeta(n-1) \label{flux-emission-eq}
\end{equation}
where $f = 1$ for bosons and $f=1-2^{2-n}$ for fermions.
For our setup, the ratio of the emitted flux into a single brane
field over a single bulk field is
\begin{equation}
\frac{d N_{\rm brane}/d t}{d N_{\rm bulk}/d t} \simeq 2.2 \,
\frac{c_{\rm brane} \, \Gamma_{\rm brane}}{c_{\rm bulk} \, \Gamma_{\rm bulk}} \; .
\end{equation}
The numerical factor $2.2$ corresponds to a $(D,n) = (7,5)$.
Varying the dimensionality $7 \le D \le 11$ and 
$5 \le n \le D-2$ one finds the numerical factor ranges from 
$1.4$ to $3.5$.  Hence, our flux estimates should be valid to 
within a factor of $1.5$ for a wide range of models.

The direct emission rate is now straightforward to calculate.  For a black 
hole in $D=7$ dimensions with gauge/Higgs fields in $n=5$ dimensions, 
we simply use Eq.~(\ref{power-suppression-eq}) and the degrees of freedom 
per species
\begin{equation}
\begin{array}{c}
c_q = 72; \;\; c_\ell = 24; \;\; c_g = 24; \;\; c_{\gamma} =3; \;\; c_h = 1; \\
c_{W^\pm,Z} = 9; \;\; c_{G^\pm,G^0} = 3; 
\end{array}
\end{equation}
to obtain the following relative direct particle emission rates
for a black hole whose horizon size is smaller than the separation
distance between quarks and leptons.  With our formalism there is no sum 
over the KK states of five dimensional fields; this is fully accounted
for in the particle multiplicities and the $n$-dimensional flux emission 
rate (\ref{flux-emission-eq}). 
For the neutrino multiplicity, both left- and 
right-handed neutrinos were included.  In five dimensions each gauge 
boson has three physical polarizations, and the bulk Goldstone bosons 
can be absorbed in the $W^\pm,Z$ multiplicity 
$c_{W^\pm,Z}^{\rm eff} = c_{W^\pm,Z} + c_{G^\pm,G^0} \Gamma_0/\Gamma_1  
\simeq 21$.  Assuming a black hole is formed 
on the quark or lepton brane and does not wander off
(this is unlikely for both energetic \cite{EHM} and kinematic
reasons), we find 
\begin{center}
\begin{tabular}{r|cccccccccccc}
                     & & $q$ & : & $\ell$ & : & $g$ & : & $W^\pm,Z$ & : & 
    $h$ & : & $\gamma$ \\ \hline
\mbox{quark brane}  & & 108 & : &    0   & : &  8  & : &   $7.1$    & : &  
    $1.4$  & : & 1 \\
\mbox{lepton brane} & &   0 & : &   36   & : &  8  & : &   $7.1$    & : &  
    $1.4$  & : & 1
\end{tabular}
\end{center}
Accounting for gauge boson decay and top quark decay, then
associating each quark or gluon as a single jet, we can estimate the final
jet to charged lepton to neutrino (plus graviton) to photon ratio,
\begin{center}
\begin{tabular}{r|cccccccc}
                    & & jets & : & $\ell^\pm$ & : & invisible & : & $\gamma$ 
\\ \hline
\mbox{quark brane}  & &  113 & : &      8     & : &   $7.5$   & : & 1 \\
\mbox{lepton brane} & &   17 & : &     20     & : &     19    & : & 1 
\end{tabular} 
\end{center}

This leads to very striking consequences for black hole production
and evaporation.  At a hadron collider, the types of partonic collisions 
include quark-on-quark, quark-on-gluon, or gluon-on-gluon.  
The first two types of partonic collisions produce a black hole 
on the quark brane that evaporates into the final state particles 
as specified above.
Gluon fusion, however, could naively create a black hole anywhere
in the extra dimension(s) in which gluons can propagate.  
The evaporation spectrum is likely to be quite unusual, dominantly 
into bulk modes only. However, gluon fusion 
is not expected to be a dominant production process due to the
relatively low gluon luminosity for producing a very heavy object.
For illustration, consider a VLHC with a center-of-mass energy of 
200 TeV. The production cross section for $\Mbh=50$ TeV for $D=11$
from $q\overline{q}$ annihilation is about $15$--$110$ fb for 
$M_*=20$--$50$ TeV, leading to a few thousand events annually. 
We have argued that $M_*$ must be at least in the tens of TeV region
in our setup, and so the LHC and lepton colliders are not energetic 
enough to produce black holes, although they could find the KK 
excitations of gauge/Higgs bosons.
(Collisions of heavy nuclei might also be of interest \cite{heavynuclei}.)

Black holes could also be produced through high energy cosmic-ray 
collisions.  If the primary incident particle is a proton, one would 
expect hadronic showers to dominate; conversely if the primary is 
a neutrino, there would be a large fraction of energetic charged leptons.
Black hole production from neutrino collisions, however, proceed only 
through $\nu g$ interactions
that are significant only for lighter black holes and at higher energies.
Typically, the cross section for $M_{bh}=50$ TeV for $D=11$ is about 
$0.1$--$1$ pb at $E_\nu \simeq 10^{19}$ eV.

In summary, we have found that in models with large extra dimensions 
and separated fermions the evaporation spectrum of black holes
significantly changed once the black hole horizon is smaller than 
the separation distance between quarks and leptons.
For example, our estimates suggest that black holes produced at 
hadron colliders in this regime will emit a large multiplicity of
jets over leptons or photons.
While we have concentrated on the broad characteristics
of black hole evaporation with separated fermions, there are 
inevitably several improvements that could be done to
further refine our predictions; we list a few of the more important ones:
Throughout this analysis we have neglected the angular momentum 
of the black hole.  In general, as argued in \cite{GT,KH}, black holes 
formed in colliders or cosmic-ray collisions are very likely to have 
significant angular momentum, and so accounting for the spin-down phase 
is quite important to accurately predict the energy and angular distributions 
of the black hole emission.  Another improvement is to 
calculate the energy-dependent greybody factors for arbitrary 
spacetime dimensionality.  Another is to account for the backreaction 
of high energy particle emission on the black hole geometry; in particular, 
accounting for the deviations from a thermal spectrum.  Finally, new stringy 
dynamics could alter the very final evaporation spectrum \cite{DEandother}
depending on the relative hierarchy between $M_*$ and the string scale.
With evidence for black holes in hand, we may well learn the spacetime
structure and standard model field geography through a detailed 
analysis of particle counting and energy spectra resulting 
from black hole evaporation.

\begin{acknowledgments}

We thank J. Hewett, N. Kaloper, and S. Thomas for encouragement and 
discussions.  GDK thanks the SLAC theory group and ITP Santa Barbara 
for their generous hospitality where this work was begun.
This work was supported in part by the U.S. Department of Energy 
under contract DE-FG02-95ER40896 and in part by the Wisconsin 
Alumni Research Foundation.

\end{acknowledgments}


\end{document}